\documentclass[journal]{IEEEtran}
\usepackage{graphicx}
\usepackage{amsmath,amsfonts}
\usepackage{siunitx}
\usepackage[font=footnotesize]{subfig}
\usepackage{float}
\usepackage{marginnote}
\usepackage{url}
\usepackage{tabularx}
\usepackage{booktabs}
\usepackage{gensymb}
\usepackage{flushend} 

\newcommand{\etal}{\textit{et al}.}
\newcommand{\ie}{\textit{i}.\textit{e}., }
\newcommand{\eg}{\textit{e}.\textit{g}. }
\usepackage{lipsum}

\ifCLASSINFOpdf
\else
\fi

\def\w{{\mathbf w}}
\def\th{{\mathbf \theta}}
\usepackage{color}

\begin{document}
%
\title{Estimation of Z-Thickness and XY-Anisotropy \\ of Electron Microscopy Images \\ using Gaussian Processes}
%
%
%


\author{Thanuja~D.~Ambegoda, Julien~N.~P.~Martel, Jozef~Adamcik, Matthew~Cook, Richard~H.~R.~Hahnloser 
\thanks{TD Ambegoda, JNP Martel, M Cook and RHR Hahnloser are with the Institute of Neuroinformatics, Univeristy of Zurich and ETH Zurich. J Adamcik is at the Department of Health Sciences and Technology, ETH Zurich.} 
 \thanks{Published by the Journal of Neuroinformatics and Neuroimaging (2017) Volume 2, Issue 2}
}

\maketitle

\begin{abstract}
Serial section electron microscopy (ssEM) is a widely used technique for obtaining volumetric information of biological tissues at nanometer scale. However, accurate 3D reconstructions of identified cellular structures and volumetric quantifications require precise estimates of section thickness and anisotropy (or stretching) along the $XY$ imaging plane. In fact, many image processing algorithms simply assume isotropy within the imaging plane. To ameliorate this problem, we present a method for estimating thickness and stretching of electron microscopy sections using non-parametric Bayesian regression of image statistics. We verify our thickness and stretching estimates using direct measurements obtained by atomic force microscopy (AFM) and show that our method has a lower estimation error compared to a recent indirect thickness estimation method as well as a relative $Z$ coordinate estimation method. Furthermore, we have made the first dataset of ssSEM images with directly measured section thickness values publicly available for the evaluation of indirect thickness estimation methods.
\end{abstract}


%
\IEEEpeerreviewmaketitle

\section{Introduction}
\label{sec:intro}
Electron microscopy (EM) has enabled imaging of nano-scale neuroanatomical structures such as synapses.
Serial section Scanning Electron Microscopy (ssSEM) and serial section Transmission Electron Microscopy (ssTEM) are used to inspect tissue volumes on the scale of tens to hundreds of micrometers in each dimension. Tissue sections suitable for ssEM typically have a thickness that ranges from $\SI{30}{nm}$ to $\SI{70}{nm}$. These extremely thin serial sections are cut from a resin-embedded specimen using an \textit{ultramicrotome} equipped with a diamond knife. Usually, there can be variations in thickness from one section to another (up to $20\%$) \cite{frank2008}. 
Another EM technique used to obtain volumetric image data is Focused Ion Beam Scanning Electron Microscopy (FIBSEM) which allows milling (virtual sectioning) in the order of $\SI{5}{nm} \sim \SI{10}{nm}$.
The problem of section thickness variation is also observed in FIBSEM data~\cite{knott2011}.   

EM image processing methods commonly implicitly assume isotropy of physical structures along the imaging plane \cite{funke2014}. However, sources of anisotropy (stretching) in the imaging plane include anisotropy intrinsic to the specimen (\eg structures with a preferred orientation), effects of sample handling and cutting, and imperfections in microscope calibration. We focus on the image analysis problem of determining the overall stretching without distinguishing between the sources of stretching. 

\begin{figure}[t]
  \centering
  \includegraphics[width=\linewidth]{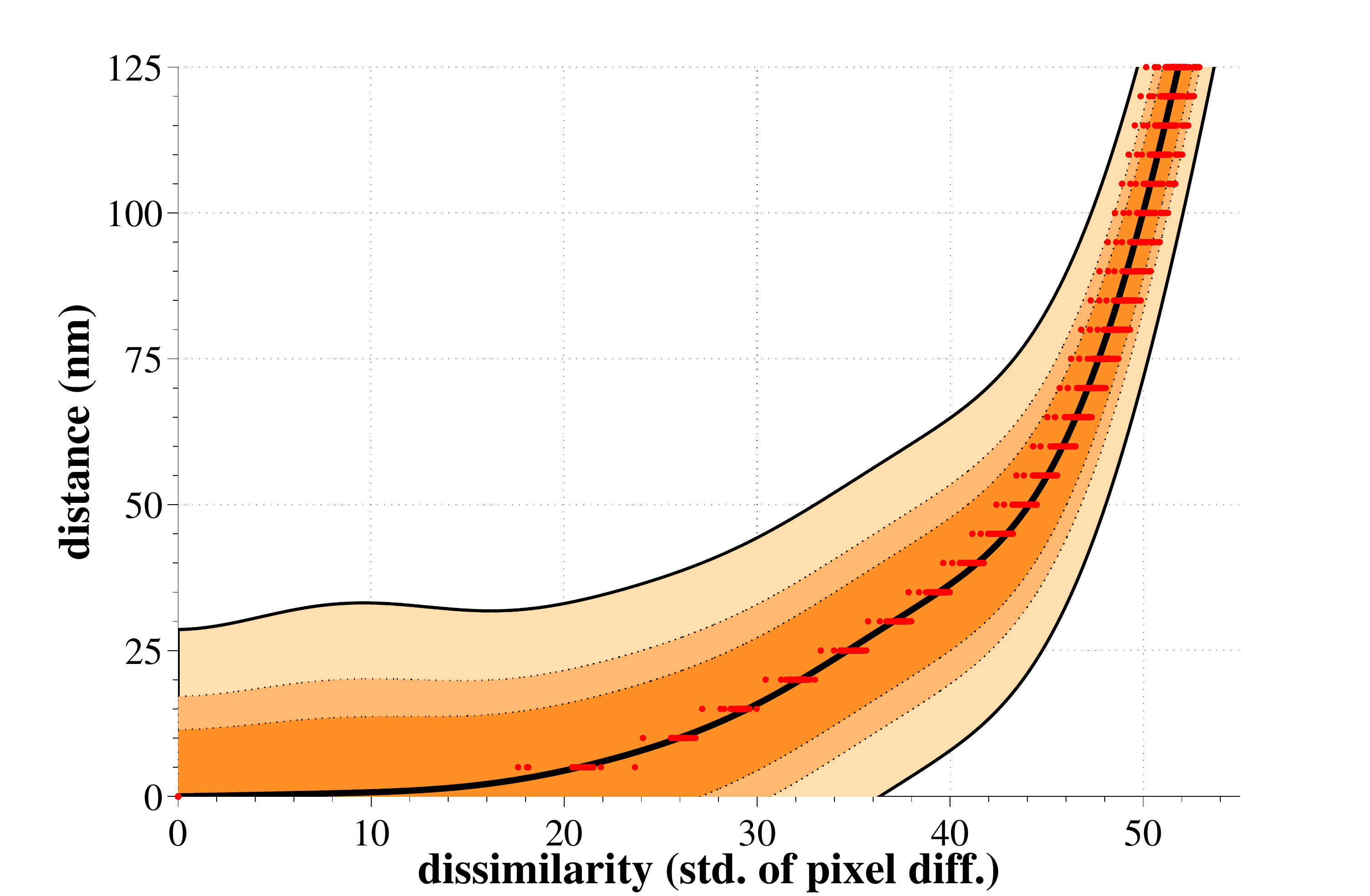}
  \caption{\small Graph of distance $D$ vs. image dissimilarity $s$, $D = f(S)$, used for the estimation of section thickness and stretching. Shown are the training data (red dots), the mean (bold line), and multiple standard deviations $\sigma$ of the Gaussian Process (GP) predictive distributions (darkest to lightest orange) $2\:\sigma(95\%)$, $3\:\sigma(99.7\%)$, $5\:\sigma$($99.\underline{9}_4\%)$.}
  \label{fig:distance_dissimilarity_curve}
\end{figure}
In this work, we address estimation of thickness and stretching by learning a function $f$ used to infer the spatial distance between pairs of sections based on image statistics. To compute predictive distributions of spatial distances for new, unseen images we use Gaussian Processes (GPs) to perform non-parametric Bayesian regression. We also use GP regressors to estimate stretching.

Section thickness estimates allow the correction of volume estimates along the $z$ axis (perpendicular to the cutting plane), which is useful for producing more accurate 3D reconstructions of imaged tissue. Furthermore, both section thickness and stretching estimates can improve the density estimation of objects such as synapses normalized per unit volume. This is particularly useful for comparing tissue volumes that underwent different experimental manipulations.

However, we note that the stretching factor alone cannot give us the original area of regions in the original sample, as it cannot distinguish tissue processing-induced stretching from any intrinsically ``stretched'' (anisotropic) nature of the original sample. Because these sources cannot be distinguished by analyzing the images alone, such absolute measurements are beyond the scope of this paper.

For any method that uses the known $XY$ resolution to model the absolute spatial distance between sections (including \cite{sporring2014} and our method), it is important to have an estimate of the anisotropy along the $XY$ plane. Such $XY$ anisotropy affects the image statistics along the two axes. If unaccounted for, the disparity of these statistics can introduce inaccuracies in the results obtained by methods that assume similar statistics along the two axes. Our solution to this problem is described in Section \ref{sec:compression}.

In order to validate the thickness estimates, we have directly measured the thickness of a set of EM sections using atomic force microscopy (AFM). We have made the validation dataset publicly available~\footnote{https://github.com/thanujadax/ssSEM\_AFM\_thickness} as a benchmark to evaluate section thickness estimation methods. 

Validation results and estimates for \textit{z}-section thickness and \textit{xy}-anisotropy for FIBSEM, ssTEM and ssSEM data sets are discussed in Section \ref{sec:results}.

The code for running the experiments described in this paper is publicly available\footnote{https://github.com/thanujadax/gpthickness}.

\section{Related work}
\label{sec:relatedwork}
In \cite{sporring2014}, the relationship between pairwise image dissimilarity and distance is computed as the average of discrete data points. The estimated thickness of a new section is interpolated from these. \cite{sporring2014} assumes that locally, images are realizations of an isotropic and rotationally invariant process. By contrast, we adopt an approach that is less affected by sample anisotropy in the $XY$ plane.
 
In \cite{hanslovsky2015} and \cite{hanslovsky2017},  the positions of the images along the $Z$ axis are iteratively corrected to seek a consistent solution in which adjacent sections have an optimal gap (or thickness) between them. The optimal solution adjusts the positions of the images such that the distance-similarity curve is maximally smooth after a fixed number of iterations.

To keep the section thickness consistent in the FIBSEM milling process \cite{boergens2013} presents a method to infer the section thickness using the intensity of the ion beam that's used for milling. Moreover, they propose to estimate the section thickness by rescaling $Z$ coordinates such that the peak of the autocorrelation along the $Z$ axis and the $X$ axis have the same Full Width at Half Maximum. However this method provides an average thickness value for all sections unlike \cite{sporring2014} and our method that estimate thickness for each section individually.

An \textit{ellipsometric} approach was suggested in \cite{peachey1958}, where the reflected color of the thin sections floating on water was used to coarsely estimate the thickness of the sections. 
This method provides a coarse estimate up to an accuracy of ${15}$ to $\SI{25}{nm}$ and the minimum thickness that can be estimated is about $\SI{50}{nm}$.
In \cite{small1968}, accidental folds in the EM sections are used to determine the thickness of that section. 
Furthermore, \cite{fiala2001} proposed the method of cylinders that use ``cylindrical'' mitochondria to get an estimate of section thickness under the assumption that cylindrical mitochondria can be found with their axis parallel to the cutting direction.


\section{Estimation of section thickness}
\label{sec:thickness}

We propose to learn a function of pairwise image dissimilarity to estimate the distance between a pair of sections. Our approach adapts the work of \cite{sporring2014} with the variation described below. What we refer to as \textit{section thickness} is the distance between a pair of adjacent sections in a series of volumetric images.

\begin{figure}[t]
  \centering
  \includegraphics[width=\linewidth]{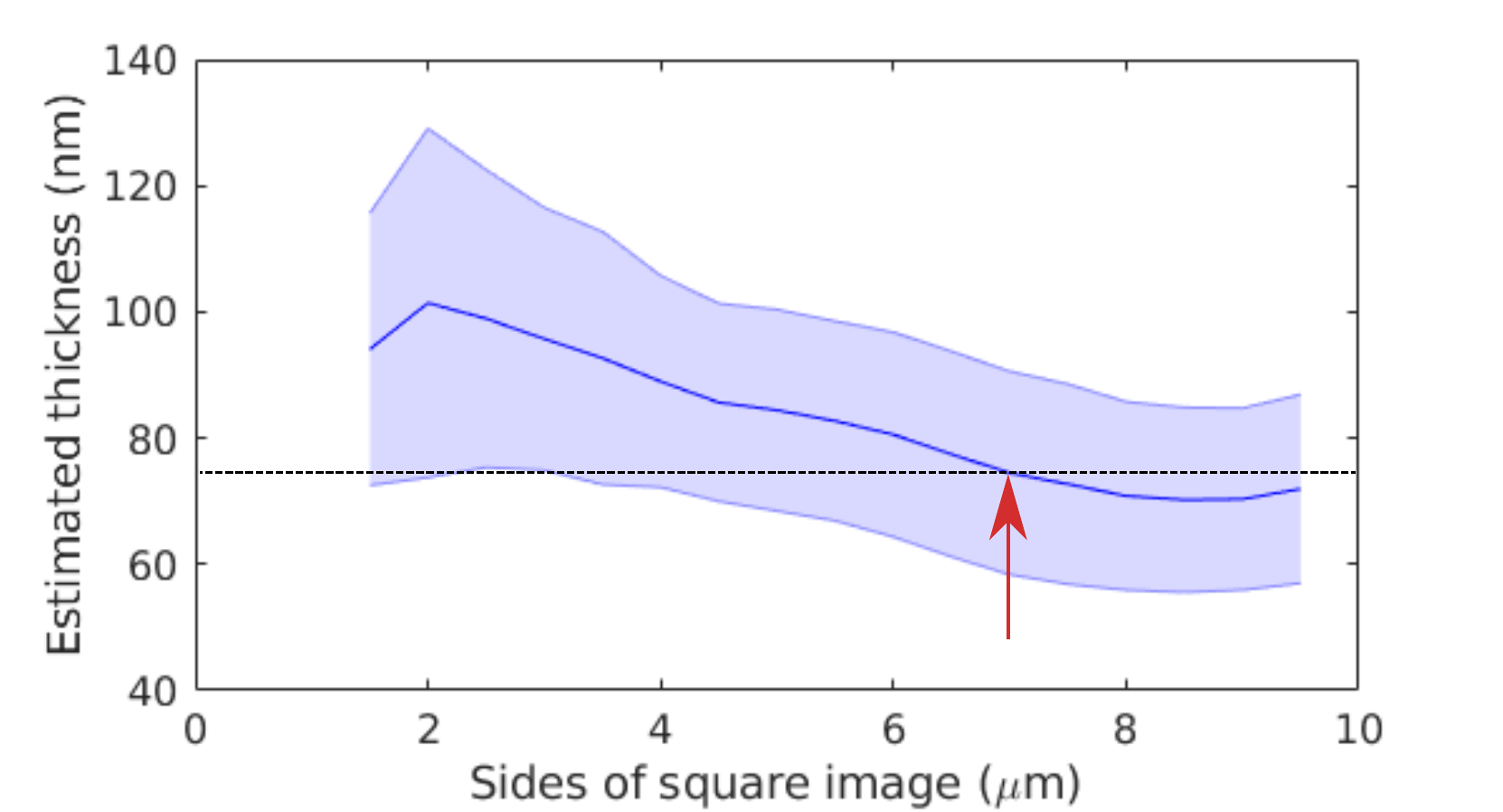}
  \caption{\small Mean thickness estimate (blue line) and and stadard deviation (Shaded area) as a function of the image size used for thickness estimation. We used  $20$ images from subvolume $(1)$ of the validation dataset. The arrow points to the image size ($\SI{7}{} \times \SI{7}{\micro\metre}$) where the average thickness estimate is equal to the average AFM thickness measurement ($74.35 \pm \SI{2.64}{\nano\metre}$) for the same dataset which is indicated by the dotted line.}
  \label{fig:thickness_vs_imgsize}
\end{figure}

We assume that local structures in the images vary smoothly in all directions at a spatial scale larger than the section thickness. Hence, the dissimilarity $S_{I_A,I_B}$ between two parallel images $I_A$ and $I_B$ only depends on the spatial distance $D_{A,B}$ between them. 
To learn the variation of image dissimilarity as a function of the distance, we extract images at known distances along the $X$ and $Y$ axes of the imaging plane: $D_{A,B} = f(S_{I_A,I_B})$.
This can be done by generating two equally sized image patches $A$ and $B$ from any original image $I$ which are a distance $D_{A,B}=n \times \Delta x$ away from each other. Here $\Delta x$ is the length of a rectangular pixel along the $X$ axis and $n$ is the number of pixels. Image patch $A$ is centered on pixel coordinates $(x_i,y_i)$, and image patch $B$ is centered on pixel coordinates $(x_i + n \times \Delta x,y_i)$.  
We observed that patches smaller than $\SI{7}{} \times \SI{7}{\micro\metre}$ tend to have problems caused by sample inherent anisotropy (e.g. elongated mitochondria, membranes accidentally having similar orientations). The variation of thickness estimates with image size for subvolume $(1)$ of the validation dataset is plotted in Figure~\ref{fig:thickness_vs_imgsize}. 
The shape of the extracted image patches has no effect on the learned statistics.

As dissimilarity measure $S_{I_A,I_B}$, we use the \textit{standard deviation of pixel-wise intensity differences (SDI)} defined in Equation~\eqref{eq:dissimilarity}, similar to \cite{sporring2014}.
\begin{equation}
\label{eq:dissimilarity}
	S_{I_A,I_B} = \sqrt{ \frac{1}{N} \sum_{x,y} { ( I_{A_{x,y}} - I_{B_{x,y}} )^2}  }
\end{equation}

We learn two separate distance-dissimilarity functions $f_x(S)$ and $f_y(S)$ as described in Section \ref{sec:gp} using images generated along the two axes $X$ and $Y$.
After estimating the relative stretching $\gamma$ between the two axes (Section \ref{sec:compression}), we use one of these functions to estimate section thickness depending on the value of $\gamma$. Since samples could be compressed in one direction relative to the other due to effects of tissue handling/cutting, we recommend using the distance-dissimilarity function corresponding to the lesser compressed axis for estimation of section thickness. 
\begin{figure}[t]
  \centering
  \includegraphics[width=\linewidth]{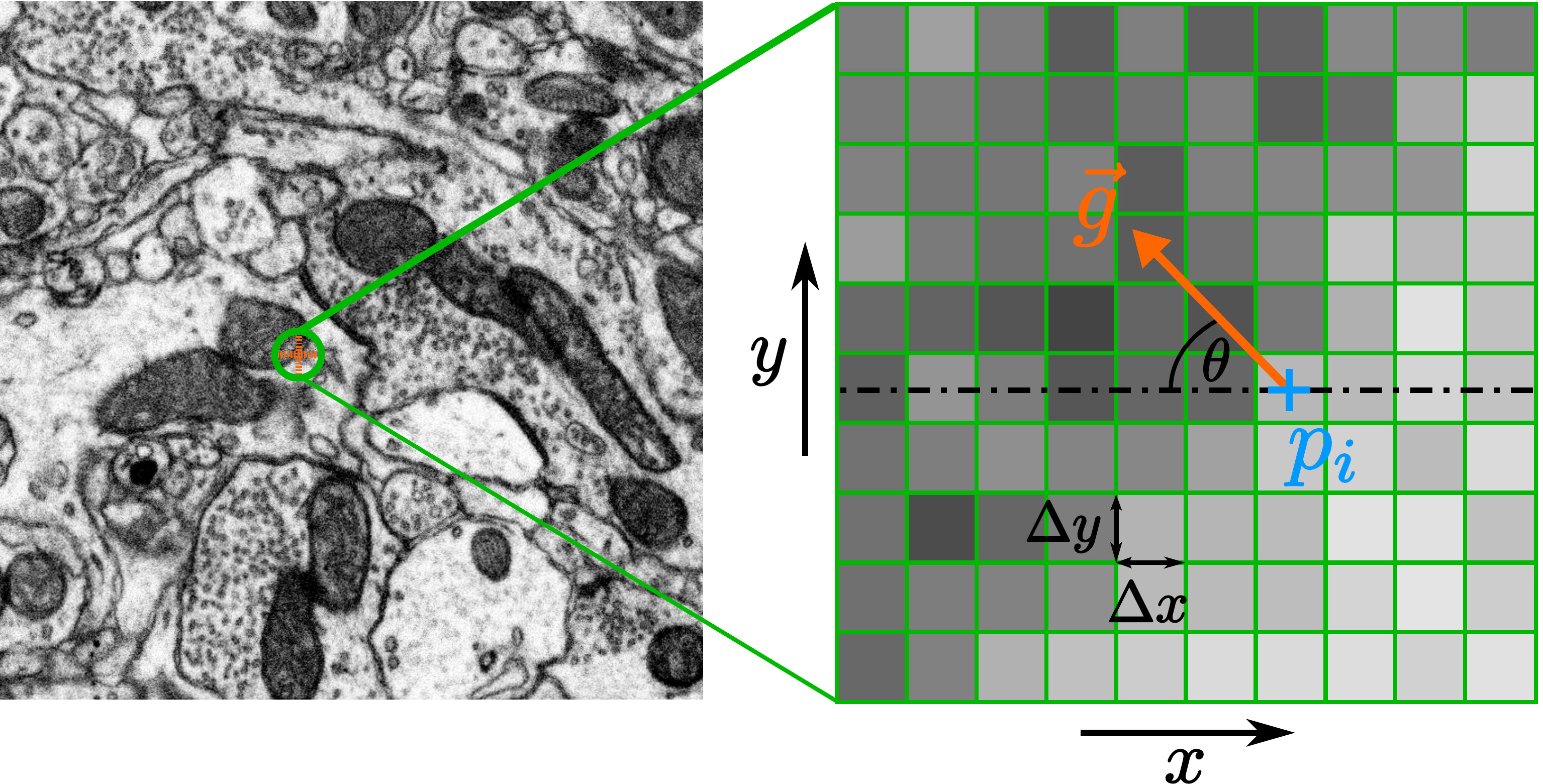}
  \caption{\small Left: FIBSEM image of $700 \times 700$ pixels. Right: An image patch with gradient $\vec{g}$ located at $p_i$ forming an angle $\theta$ w.r.t to the $X$ axis.}
    \label{fig:compression}
\end{figure}

\section{Non-parametric Bayesian estimation using Gaussian process regression}
\label{sec:gp}

We aim to learn from data the distance $D$ of two images as a function $f$ of image dissimilarity $S$ between pairs of images:
\begin{equation}
	D = f(S)
    \label{eq:fs}
\end{equation}
Given many image pairs, a training dataset consisting of $N$ data points $\{(D_i,S_i)\}_{i\in\{1,...,N\}}$ in the distance-dissimilarity plane is created. This general supervised learning framework in which we estimate the function $f$ that best fits these data points is commonly known as regression. The use of a regression model to infer an output (displacement in our case) given an input (our dissimilarity) is usually referred to as prediction.

In regression analysis, a common method to learn $f$ is to assume a specific form $f_\w$ parameterized by a vector $\w$. Then, the regression problem can be formulated as finding the best set of parameters that minimizes a sample loss $L_i$, for all pairs of outputs $D_i$ and inputs $S_i$.
As an example, least squares regression specifies $L_i = \Vert D_i - f_\w(S_i) \Vert_2^2$ and finds $\w^*$, an optimal set of weights such that $\w^* = \text{arg min}_{\w}\text{ }\sum_i L_i$. 

Here we formulate the regression problem in a Bayesian framework that aims to infer the posterior distribution of the parameters $p(\w|D,S)$, given a prior for their distribution $p(\w)$ and a likelihood coming from the data $p(D|S,\w)$, $D=f_\w(S)+\epsilon$, where $\epsilon$ is a noise model.
In this view, the mode of the posterior $p(\w|\{D_i,S_i\}_i)$ corresponds to the most likely solution for the regressor, and the standard deviation of the posterior corresponds to the uncertainty.

An intrinsic limitation of parametric regression is the need to explicitly specify $f_\w$: in many practical problems, this function is a-priori unknown and one might prefer not to make strong assumptions about it. 

We formulate the inference problem in function space \cite{gpbook2006} using Gaussian Process (GP) regression \cite{gpprediction1999}, where a GP is set of random variables for which any finite subset has a joint Gaussian distribution. A GP defines a probability distribution over functions that allows inference in the space of functions. 
The GP is completely specified by a mean function $m(S)$ and a covariance function $k(S,S')$ reflecting the mean and covariance of the process $f$, formally: $m(S)=\mathbb{E}\big[f(S)\big]$ 
and, $k(S,S')=\mathbb{E}\big[\big(f(S)-m(S)\big)\big(f(S')-m(S')\big)\big]$, where $\mathbb{E}$ denotes the expectation w.r.t $f$. The unknown function $f(S)$ can be seen as a realization of the Gaussian Process:
\begin{equation}
	f(S)\sim \mathcal{GP}\big(m(S),k(S,S')\big)
\end{equation}
To perform regression using a Gaussian Process \cite{gpregression1996}, only the form of the functions $m$ and $k$ have to be specified. These functions do not induce a particular form for $f$: rather the covariance function can be seen as specifying a prior over the function space. 
Also, note that eventual hyperparameters for the mean and covariance function can be learned from data. 

For our model, we choose the covariance $k(S,S')$ function to be a squared exponential (SE): 
\begin{equation}
	k(S,S')=\sigma^2\exp\Big(-\frac{1}{2\:l^2}\:(S-S')^2\Big)
\end{equation}
Intuitively, an SE covariance is a smoothness prior on the functions determined by the length-scale $l$ and signal standard deviation parameter $\sigma$. For the distance measure SDI we choose a function of the form $m(S)=a\:S^b$ as the mean function, because the SDI empirically shows a power law increase starting from $(0,0)$.
\begin{equation}
\label{eq:powerfunction1}
	m(S)=a\:S^b
\end{equation} 
%

The set of hyperparameters $\mathbf{\th}=(\sigma,l,a,b)^T$ can be seen as hyperpriors that guide the optimization but are not restrictive. Both hyperpriors $p(a)=\mathcal{N}(\mu_a,\sigma_a)$ and $p(b)=\mathcal{N}(\mu_b,\sigma_b)$ are Gaussian distributions whose means $(\mu_a,\mu_b)$ and standard deviations $(\sigma_a,\sigma_b)$ are determined using a standard non-linear regression of a function $D'=a\cdot S'^{b}$ using a Levenberg-Marquardt algorithm. For the distance measure SDI we empirically found that the length scale parameter $l=10$ and the signal variance parameter $\sigma = 1$ allow the GP to model the desired level of smoothness and robustness to noise by visual inspection of resulting distance-similarity plots.

To perform the GP regression itself, we compute the marginal likelihood. We then produce a predictive distribution for the output $D$ at each test input location $S$\@. Figure~\ref{fig:distance_dissimilarity_curve} shows the mean and variance of these predictive distributions.

\section{Estimation of stretching}
\label{sec:compression}
%
\begin{figure}[t]
 	\centering
    \subfloat[]{\includegraphics[width=1\linewidth]{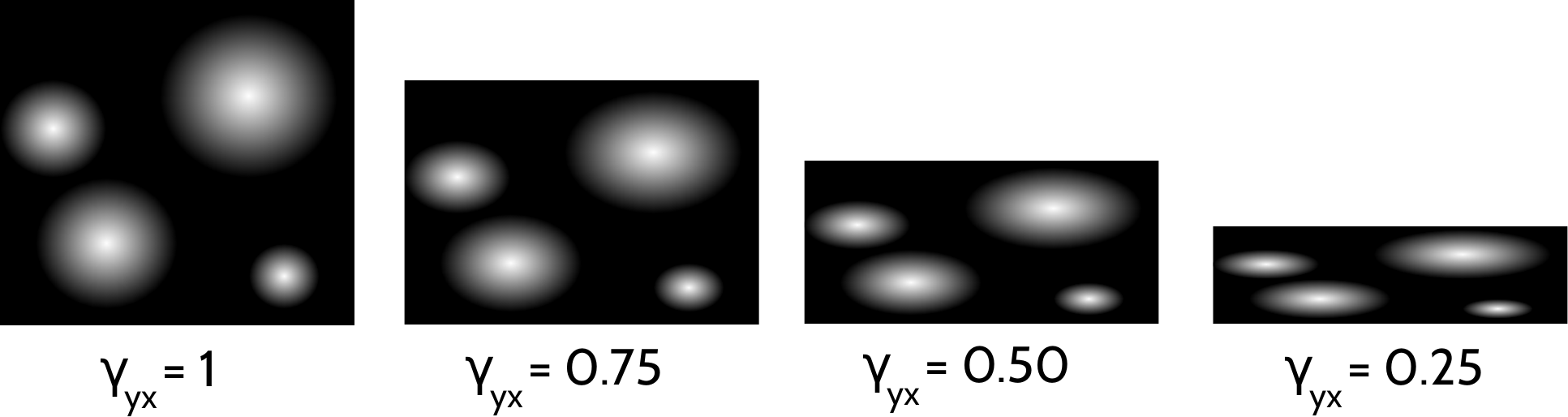}\label{subfig:radial_grad}}
	\hfil
    \subfloat[]{\includegraphics[width=0.48\linewidth]{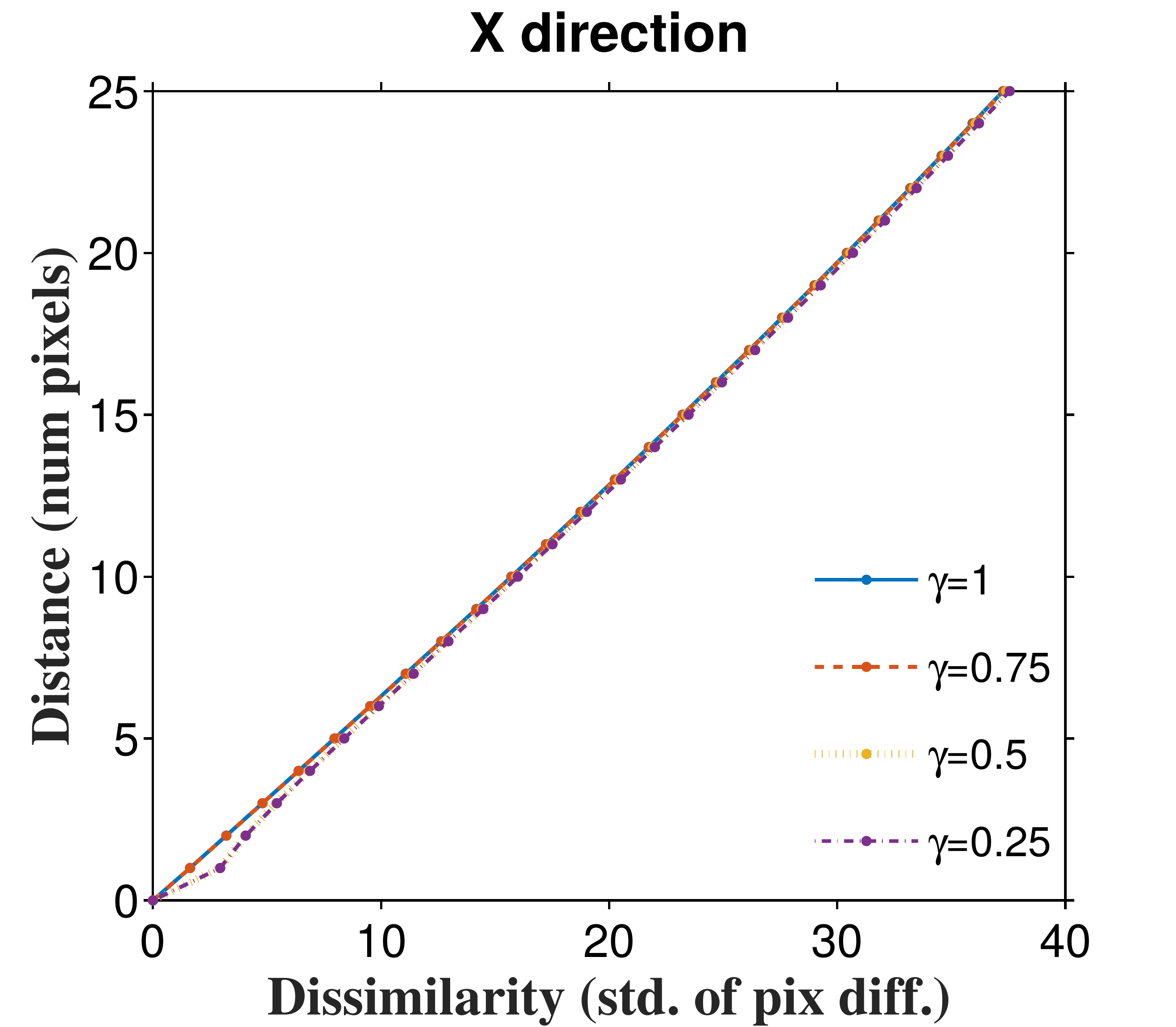}\label{subfig:alongX}}
    \hfil
    \subfloat[]{\includegraphics[width=0.48\linewidth]{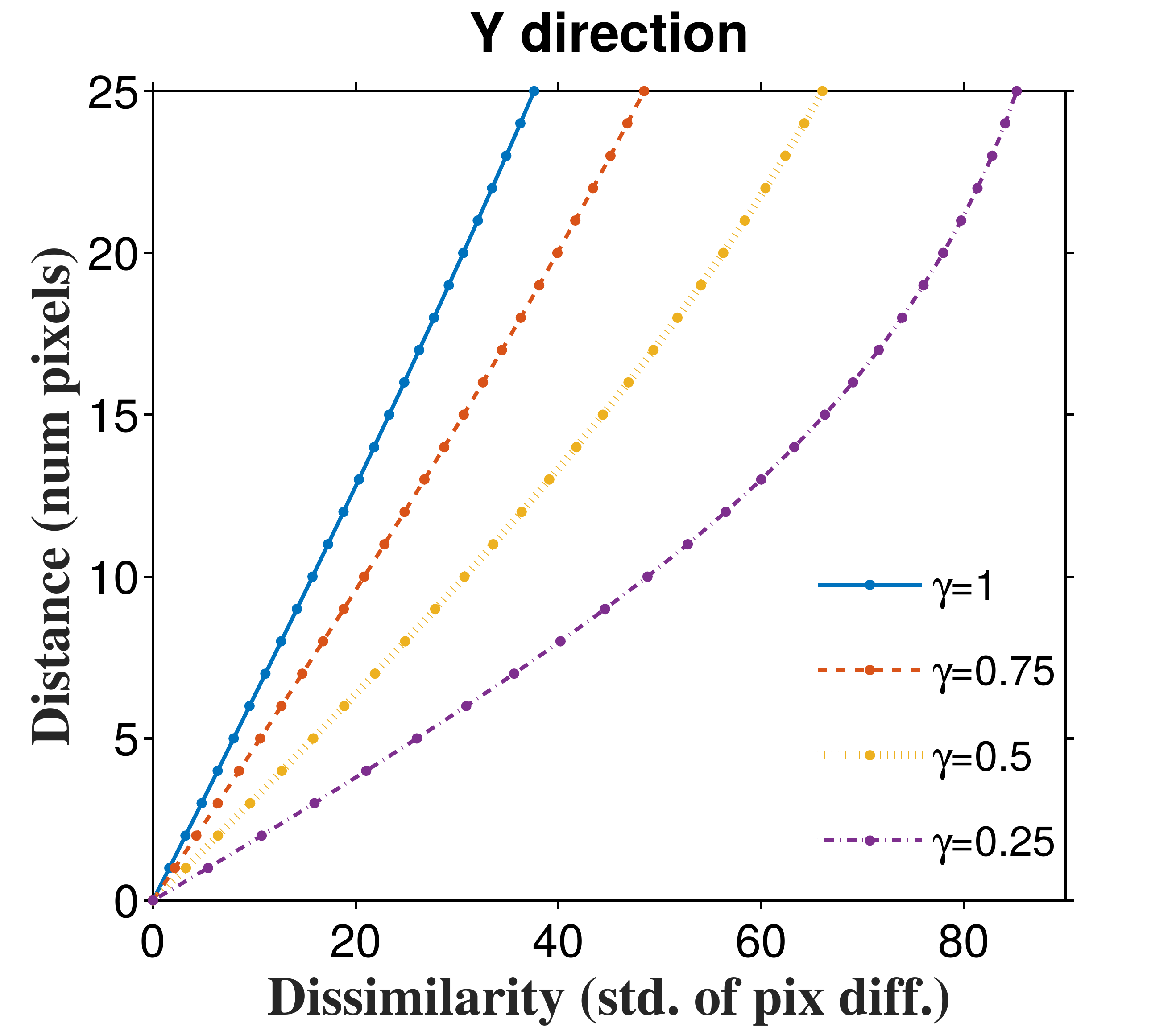}\label{subfig:alongY}}
  \caption{\small
  			\protect\subref{subfig:radial_grad} shows four images with an artificially generated pattern featuring locally radial gradients. From left to right the images underwent a simulated linear stretching (in this case, compression) along the $y$ axis while the $X$ axis remained unchanged. The legend specifies the normalized size $w\times h$ of the image.  
  			\protect\subref{subfig:alongX} shows the distance-dissimilarity plots for shifted versions of the radial gradient pattern along the horizontal axis
        and \protect\subref{subfig:alongY} along the vertical axis.}
\label{fig:squashed_images_curves}
\end{figure}
%
The learned distance-dissimilarity function can be utilized to estimate the \textit{stretching coefficient} $\gamma_{yx}$, defined as the deviation from isotropy of the image along the $Y$ axis relative to the $X$ axis.
Consider a small image patch with pixel intensity gradient $\vec{g}$ at an angle $\theta$ relative to the $X$ axis (Figure~\ref{fig:compression}). The intensity difference $\Delta p_i$ at pixel $i$ between two image patches separated by one pixel ($\Delta x$) along the $X$ axis is given by: $\Delta p = p_i - p_i \: \Vert\vec{g}\Vert \cos \theta \: \Delta x$, where $p_i$ is the pixel intensity at pixel $i$. It follows from Equation~\eqref{eq:dissimilarity} that the dissimilarity between these image patches (ignoring boundary conditions) is:
\begin{equation}
	S(I_{x,y},I_{x+\Delta x,y}) = \Delta x \: \Vert\vec{g}\Vert \cos \theta\sqrt{\frac{1}{N}\sum_{i=1}^{N} {p^2_i} }
\label{eq:similarity_gcostheta}
\end{equation}
As shown by Equation~\eqref{eq:similarity_gcostheta} the dissimilarity is directly proportional to the local gradient of the image patch. We use this result to estimate $\gamma$ along one axis relative to the other (because stretching along one axis alters the component of the gradient along that axis).

To estimate $\gamma$ along the $Y$ axis relative to the $X$ axis (\ie $\gamma_{yx}$) we perform the following steps: First, the distance-dissimilarity function $f_x(S)$ is learned using images displaced by $n$ pixels along the $X$ axis. Then, for a pair of images separated by one pixel along the $Y$ axis (distance $\Delta y$), we calculate the dissimilarity value $S$ using Equation~\ref{eq:dissimilarity}. Using the value of $S$, we estimate the pixel distance using the regression function $f_x(S)$ learned above. 
This estimate gives $\hat{n}_{yx}$, where $\hat{\Delta y} = \hat{n}_{yx} \: \Delta x$.
This is the expected length of a pixel along the $Y$ axis using the distance-dissimilarity statistics along the $X$ axis. Therefore, $\hat{n}_{yx}$ captures the linear scaling of the $Y$ axis with respect to the $X$ axis in terms of distance-dissimilarity statistics.
The stretching coefficient $\gamma_{yx}$ of the $Y$ axis relative to the $X$ axis is defined as 
\begin{equation}
	\gamma_{yx} = \frac{\Delta y}{\hat{\Delta y}} = \frac{ \Delta y}{\hat{n}_{yx} \: \Delta x}
    = \frac{a_{yx}}{\hat{n}_{yx}}
    \label{eq:gamma}
\end{equation}
where $a_{yx}$ is the pixel aspect ratio $\Delta y / \Delta x$.
For a pixel aspect ratio of $1$, $\gamma_{yx} > 1$ implies stretching of the $Y$ axis relative to the $X$ axis.
Once the $\gamma_{yx}$ is known, we suggest to use the regressor corresponding to higher $\gamma$ (lower relative compression) as the distance-dissimilarity function for section thickness estimation. 
For instance, provided $\gamma_{yx} < 1$, the regressor $f_x(S)$ should be used because the linear compression of the $Y$ axis is potentially higher than that along the $X$ axis and therefore $f_x(S)$ will result in a more accurate thickness estimate.

However, the exact orientation of the $X$ and $Y$ axes are arbitrary. In order to find the directions of maximum and minimum stretching, $\gamma_{yx}$ has to be calculated for a range of orientations. The lowest value of $\gamma^*_{yx}$ corresponds to the pair of orthogonal axes for which $X$ has the minimum stretching along its direction. 
\section{Validation of thickness estimation using Atomic Force Microscopy}
\label{sec:afm}
Validation of EM section thickness estimation methods is limited by the unavailability of a standard data set with accurately measured thickness. We used Atomic Force Microscopy (AFM)~\cite{binning1986} to produce a dataset for validation of thickness estimates.
AFM is a scanning probe microscopy technique that can be used to measure the 3D surface profile of a section at nanometer resolution. The AFM probe is a sharp tip with a typical radius of $\SI{5}{nm} \sim \SI{50}{nm}$ that scans the surface while measuring changes in the atomic forces between the sample and the tip. AFM allows us to directly measure the thickness of ssEM sections placed on flat silicon wafers.

Bienias \etal~\cite{bienias1998},  Meli \etal~\cite{meli1998} and Garnaes \etal~\cite{garnaes2003} have carried out an uncertainty analysis for height measurements  using AFM. The uncertainty of the measurements for heights of around $\SI{200}{\nano\metre}$ is reported to be $\SI{1}{\nano\metre}$ whereas for heights below $\SI{50}{\nano\metre}$ the uncertainty is $\SI{0.5}{\nano\metre}$.

\begin{figure}[t]
  \centering
\subfloat[]{
  \includegraphics[width=\linewidth]{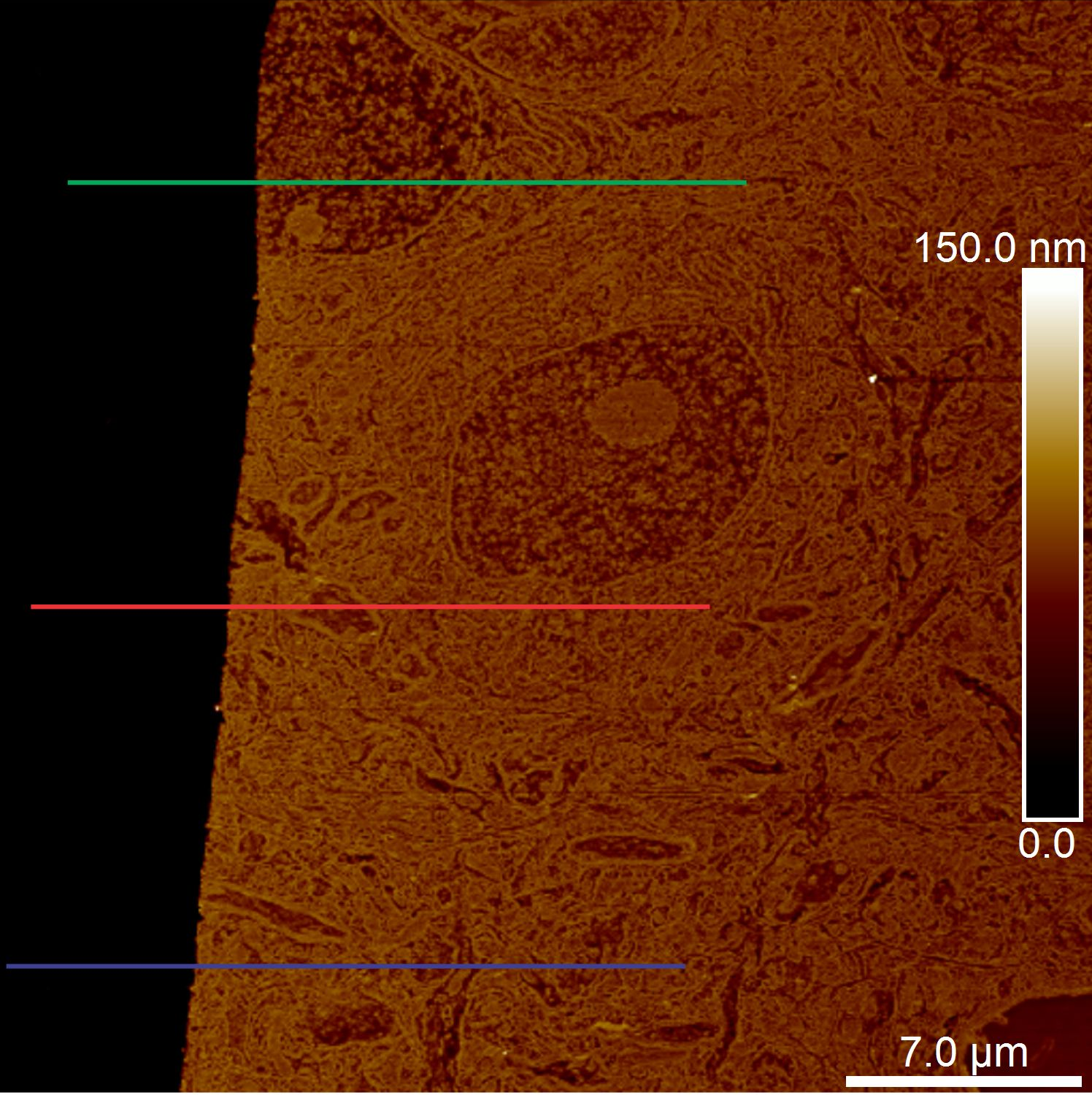}
    \label{subfig:AFMmeasurement_image} }

\hfill

\subfloat[]{
  \includegraphics[width=\linewidth]{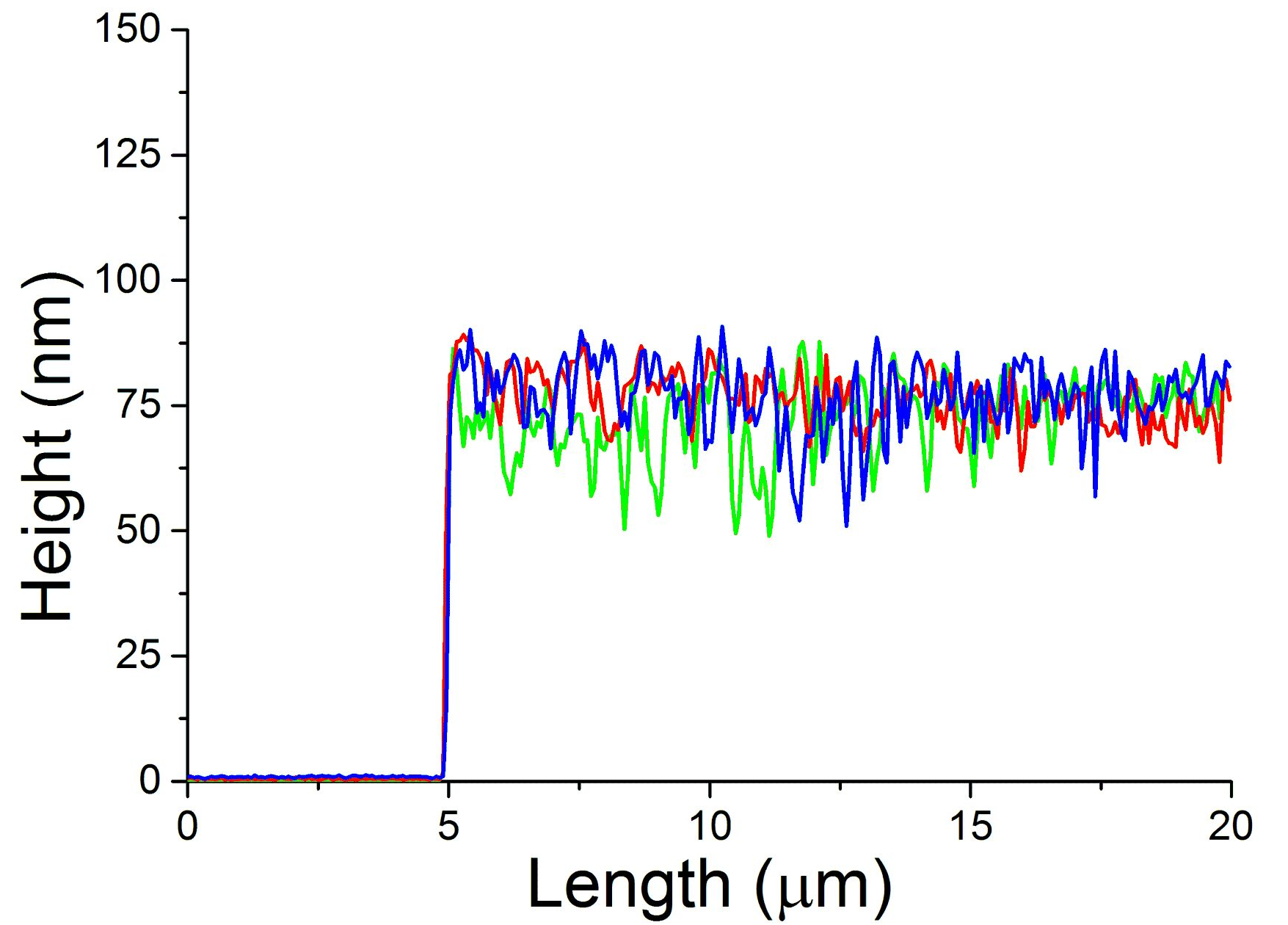}
    \label{subfig:AFMmeasurement_graph} }

\label{fig:AFMmeasurements}

\caption{\small \protect\subref{subfig:AFMmeasurement_image} is an AFM image of one section prepared for EM. The pixel intensity corresponds to the relative height difference from a reference point which is on the platform on which the section is placed. We obtained AFM measurements along three scan lines (green, red and blue) as shown in the image. This process repeated for all the sections separately.
\protect\subref{subfig:AFMmeasurement_graph} shows the measured height of the section relative to the height of the silicon wafer along each of the three colored lines in Figure~\ref{subfig:AFMmeasurement_image}. The average of these three measurements is taken as the section thickness reported by AFM.}

\end{figure}

As illustrated in Fig.\ref{subfig:AFMmeasurement_image}, thickness measurements were obtained using AFM along three distinct scan lines along each ultrathin tissue section. We measured the thickness of each section as the average distance between the surface of the silicon wafer and the surface of the EM section (Fig.\ \ref{subfig:AFMmeasurement_graph}). EM imaging (with parameters: dwell time $\SI{7}{\micro\second}$, probe current $\SI{500}{\pico\ampere}$, EHT $\SI{1.5}{\kilo\volt}$) typically mills around $\SI{10}{nm}$ of tissue. To avoid offsetting the AFM thickness measurements by EM milling, we made sure that EM imaging and AFM measurements were performed on non-overlapping regions on the sections. 

\begin{figure}[t]
  \centering
  \includegraphics[width=\linewidth]{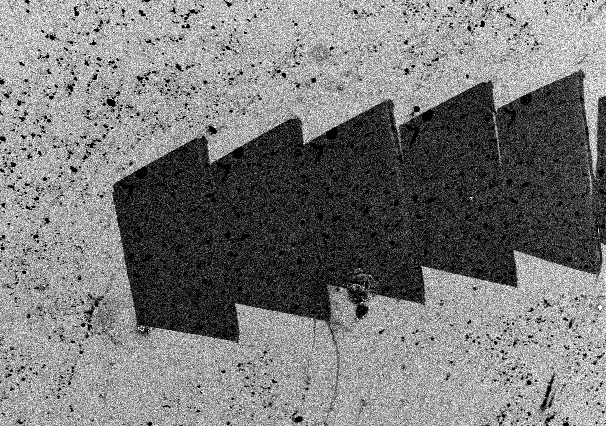}
  \caption{\small EM image of several trapezoid shaped EM sections placed on a silicon wafer. The image was obtained at a pixel resolution of $\SI{2.2}{\micro\metre} \times \SI{2.2}{\micro\metre}$.}
\label{fig:siliconWafer}
\end{figure}
\begin{figure}
\includegraphics[width=\linewidth]{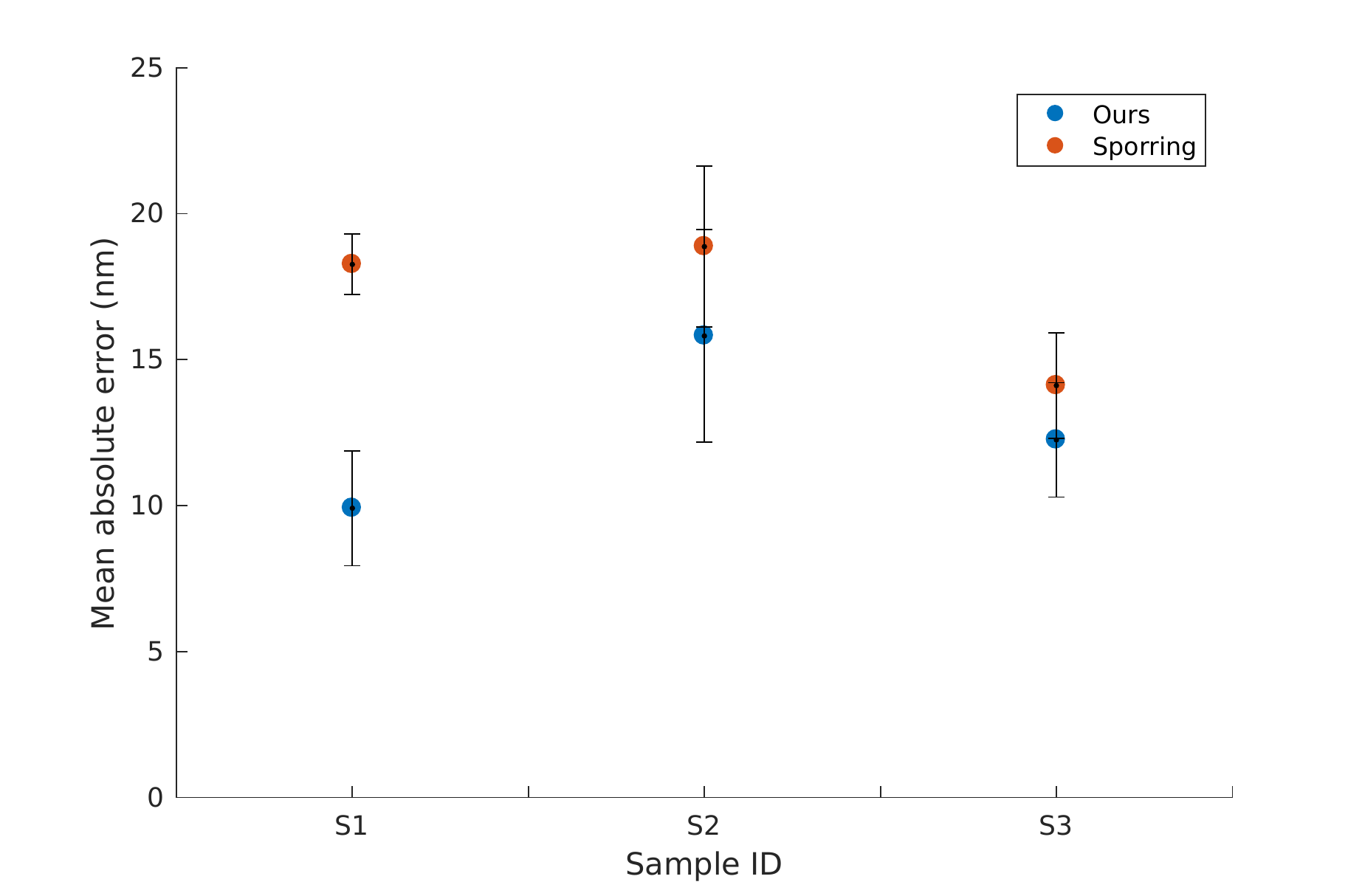}
\caption{\small Absolute estimation errors using our method and \cite{sporring2014} with respect to AFM measurements of section thickness for the three image stacks derived from the ssSEM validation dataset. Absolute estimates for sample S1 are plotted in Figure \ref{fig:AFMresults}.
Mean absolute error estimates for the three volumes using our method: $9.91 \% \pm 1.97$, $15.81 \% \pm 3.64$ and $12.25 \% \pm 1.96$; 
using Sporring \etal~\cite{sporring2014}: $18.26 \% \pm 1.04$, $18.87 \% \pm 2.75$ and $14.11 \% \pm 1.81$.}
\label{fig:afm_summary}
\end{figure}

\section{Results and discussion}
\label{sec:results}
\begin{figure}[t]
  \centering
  \subfloat[]{
  \includegraphics[width=\linewidth]{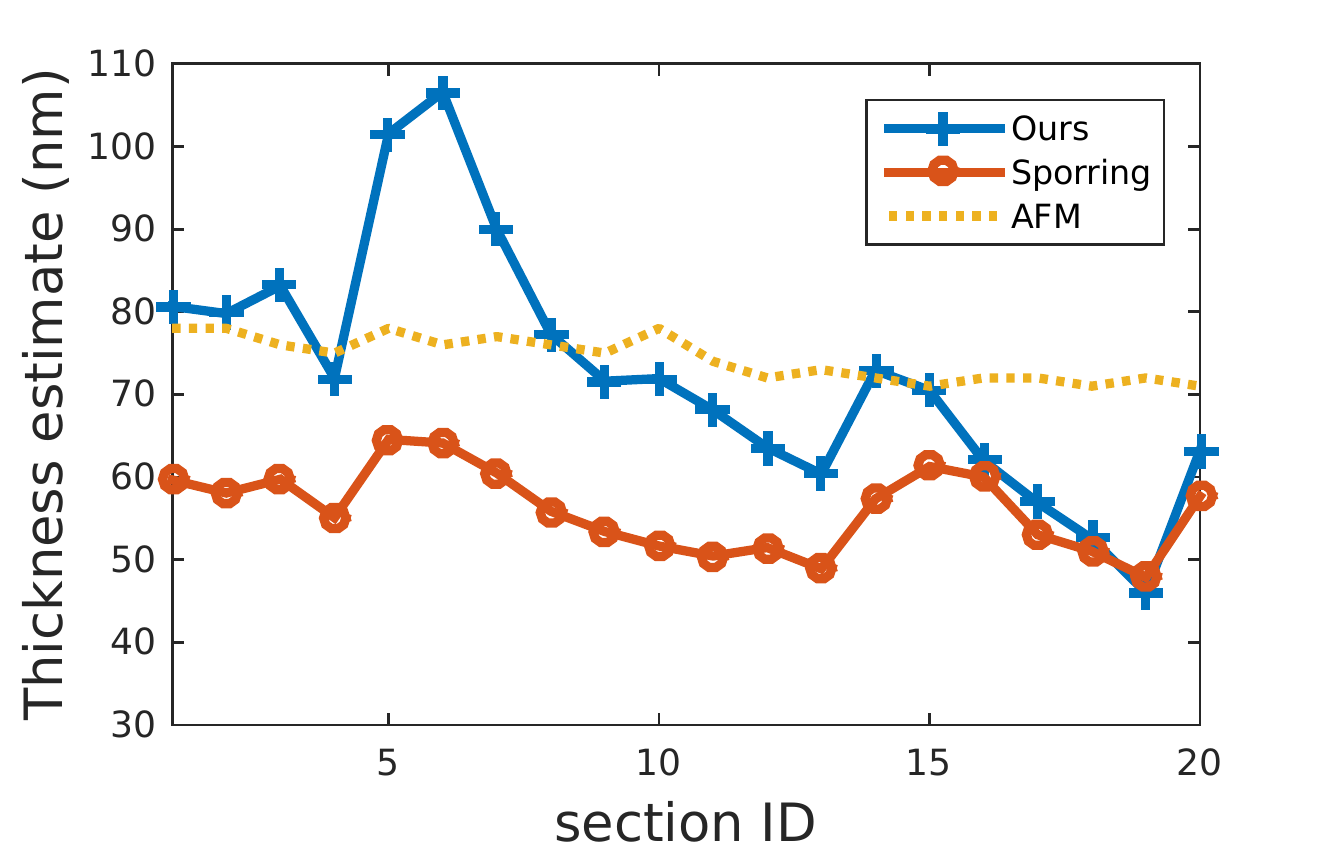}
  \label{subfig:AFMresults_1} }
  \hfill
  \subfloat[]{
  \includegraphics[width=\linewidth]{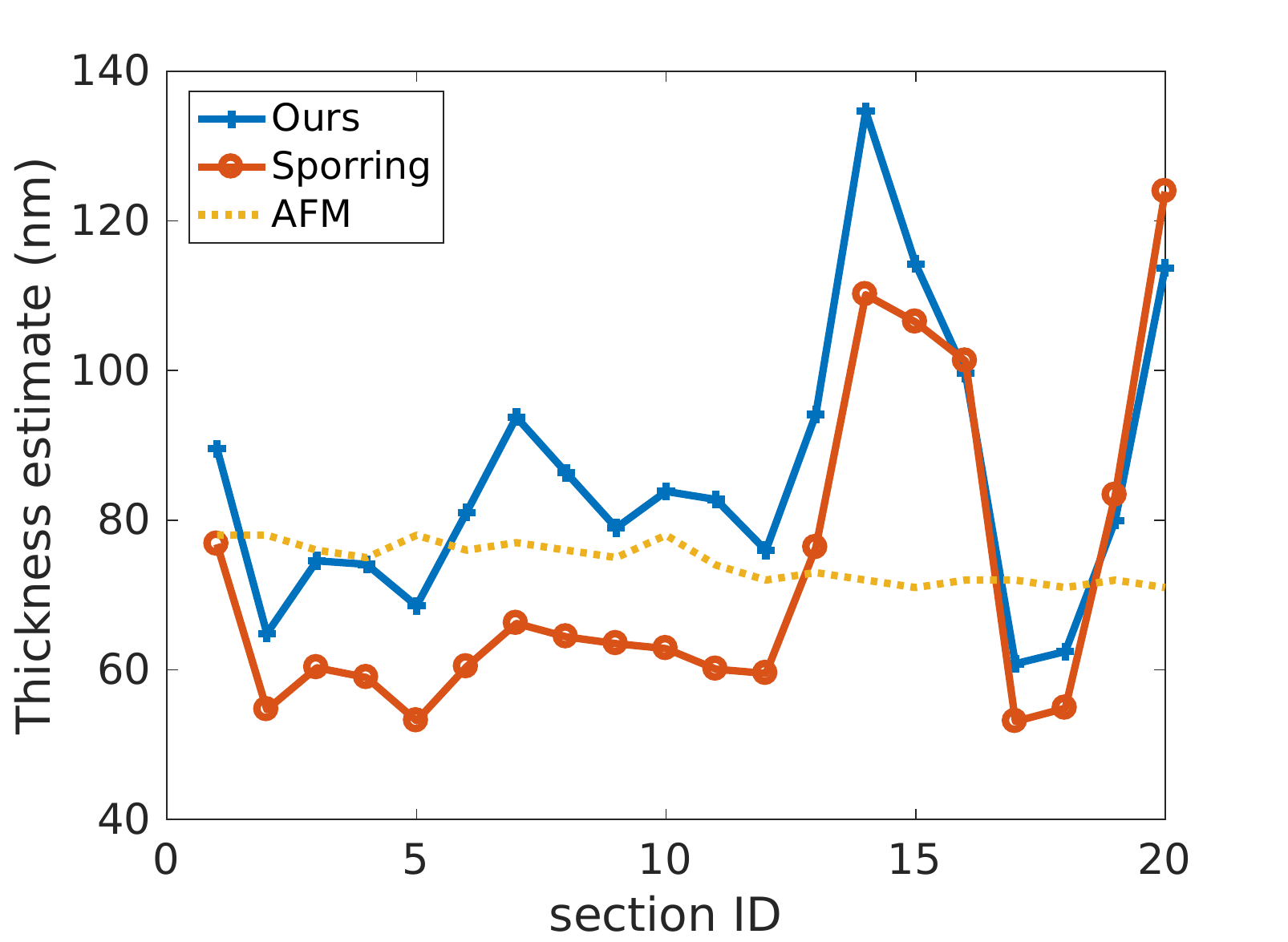}
  \label{subfig:AFMresults_2} }  
\caption{\small Validation using AFM: Comparisons of thickness estimates using our method (GP), the method of XY averaging introduced by Sporring \etal~\cite{sporring2014} and direct measurements of section thickness by atomic force microscopy (AFM), using \protect\subref{subfig:AFMresults_1} subvolume (1) of the validation dataset (image size: $\SI{9.5}{} \times \SI{9.5}{\micro\metre}$) and 
\protect\subref{subfig:AFMresults_2} subvolume (2) of the validation dataset (image size: $\SI{6.5}{} \times \SI{6.5}{\micro\metre}$).}
\label{fig:AFMresults}
\end{figure}
\begin{figure}[t]
  \centering
  \includegraphics[width=\linewidth]{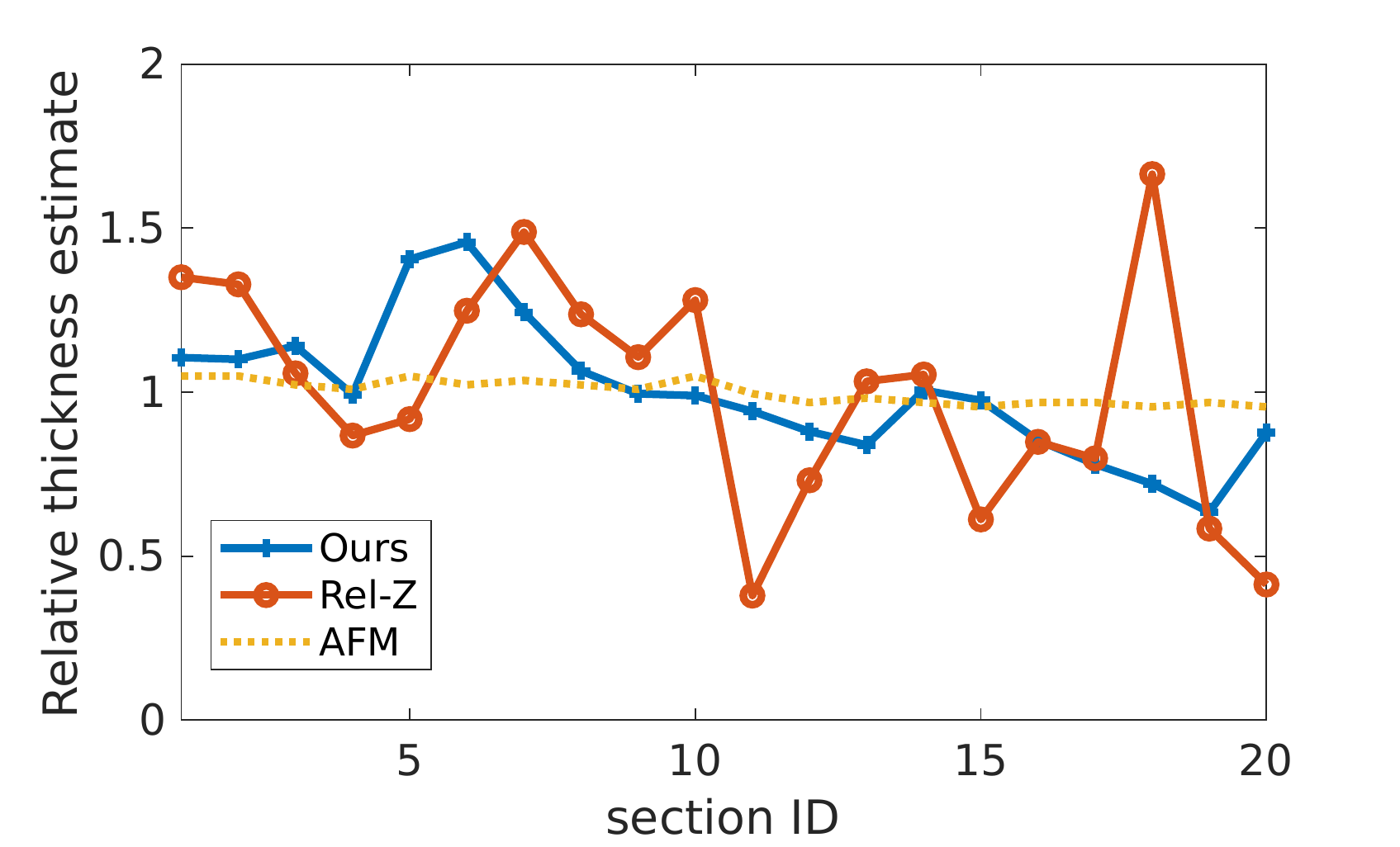}
\caption{\small  To compare our thickness estimates with relative $Z$ coordinates estimated using (Rel-Z)~\cite{hanslovsky2015}, we have produced relative thicknesses by normalizing absolute thicknesses obtained from both our method and from AFM measurements. Compared to relative thicknesses obtained wtih AFM, the mean absolute error of our method is $0.13 \pm 0.03$ whereas \cite{hanslovsky2015} resulted in $0.27 \pm 0.04$ absolute error. This plot was produced using subvolume (1) of the validation dataset.}
\label{fig:relativeThicknessComparison}
\end{figure}

\begin{figure}[t]
  \centering
\subfloat[]{
  \includegraphics[width=\linewidth]{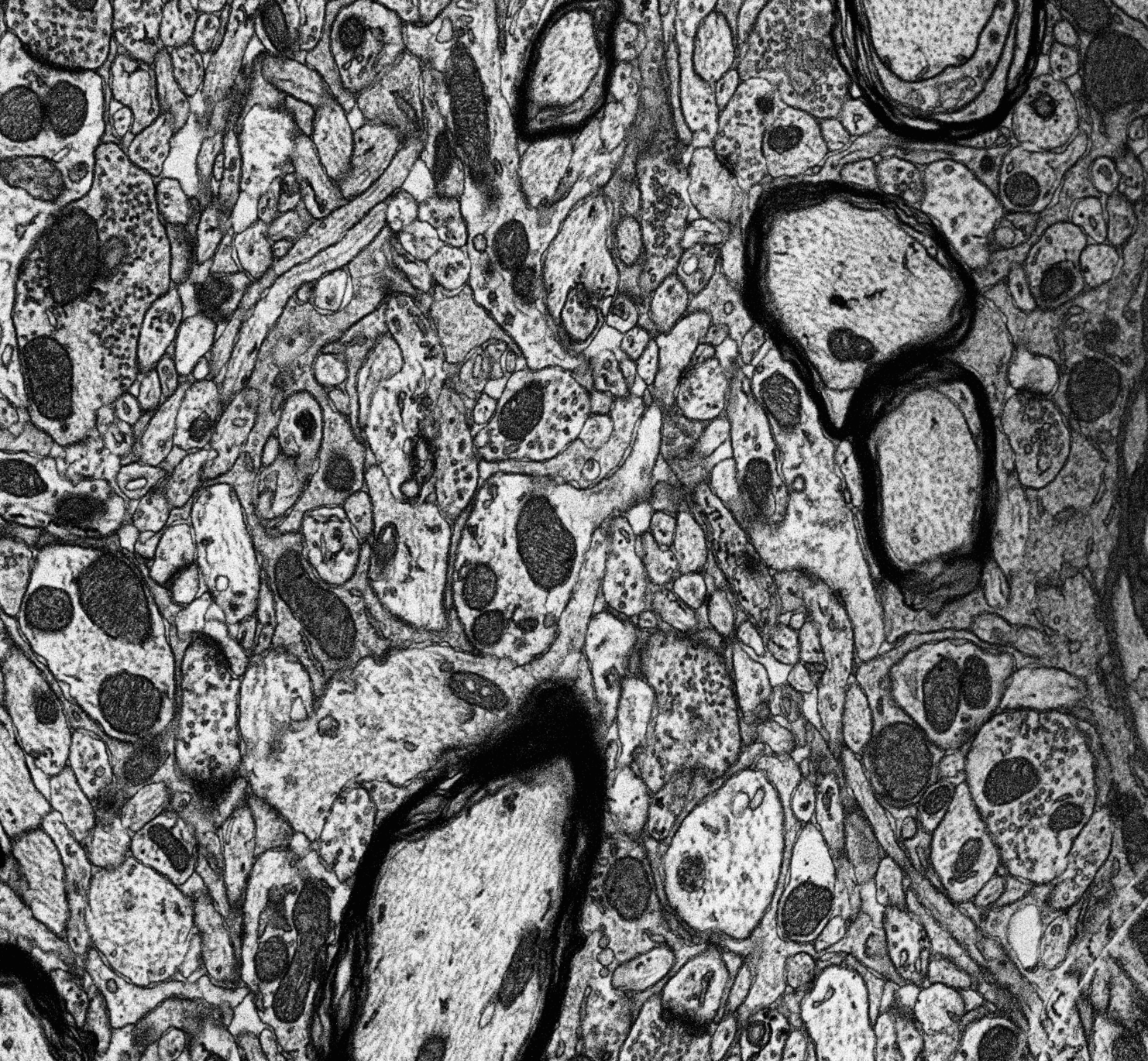}
    \label{subfig:EM_img14} }
\hfill
\subfloat[]{
  \includegraphics[width=\linewidth]{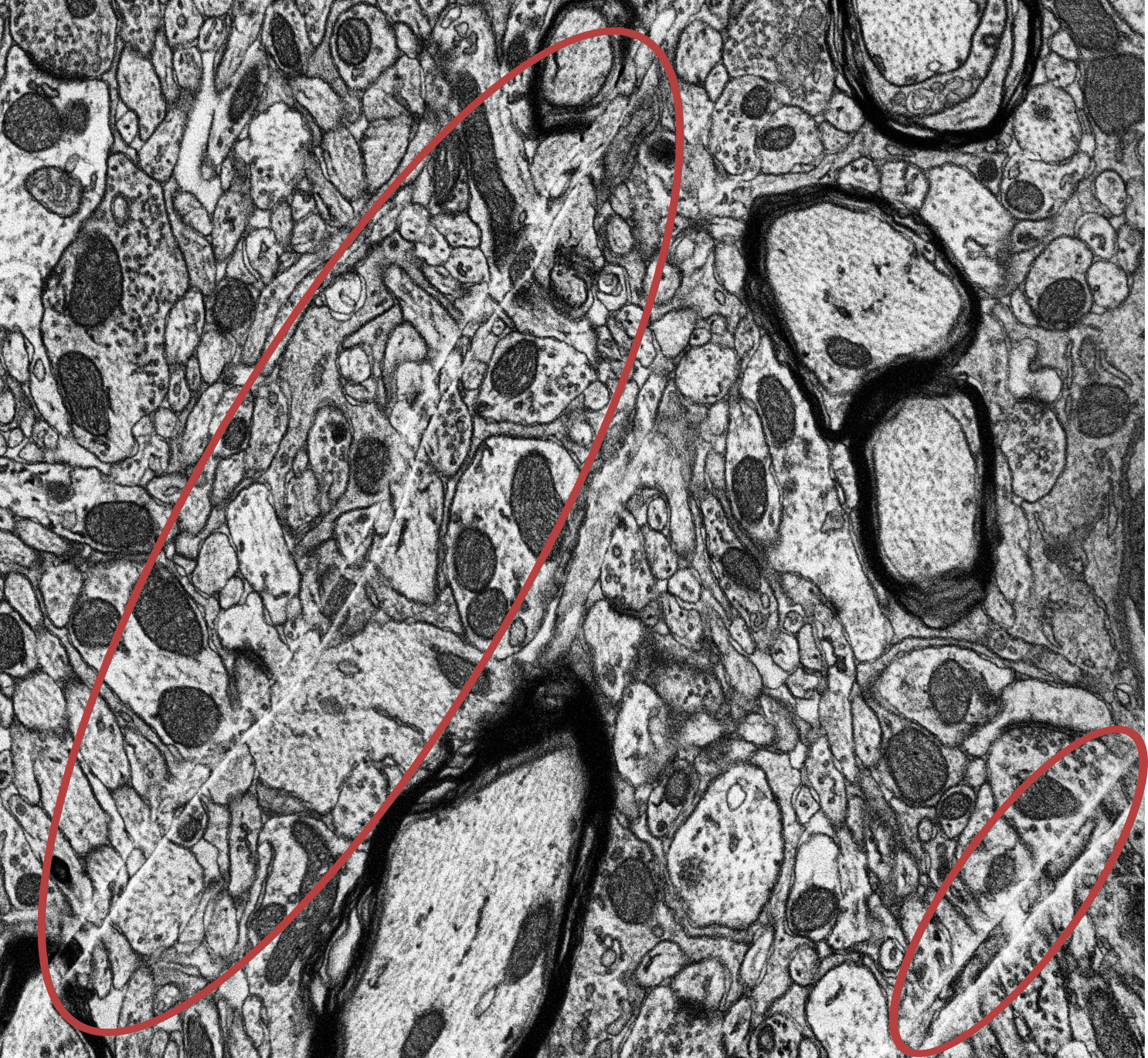}
    \label{subfig:EM_img15} }

\caption{\small Image pair corresponding to the peak thickness estimate ($sectionID=14$) of the graph shown in Fig.\,\ref{subfig:AFMresults_2}. This value is calculated based on the similarity between images \protect\subref{subfig:EM_img14} and \protect\subref{subfig:EM_img15}. Artifacts seen as white lines in image \protect\subref{subfig:EM_img15} (highlighted using red ellipses) have contributed to increase the dissimilarity between the two images, thereby resulting in an overestimate of the thickness. }

\label{fig:EM_explanation}

\end{figure}

\begin{figure}
 	\centering
    \subfloat[]{\includegraphics[width=0.9\linewidth]{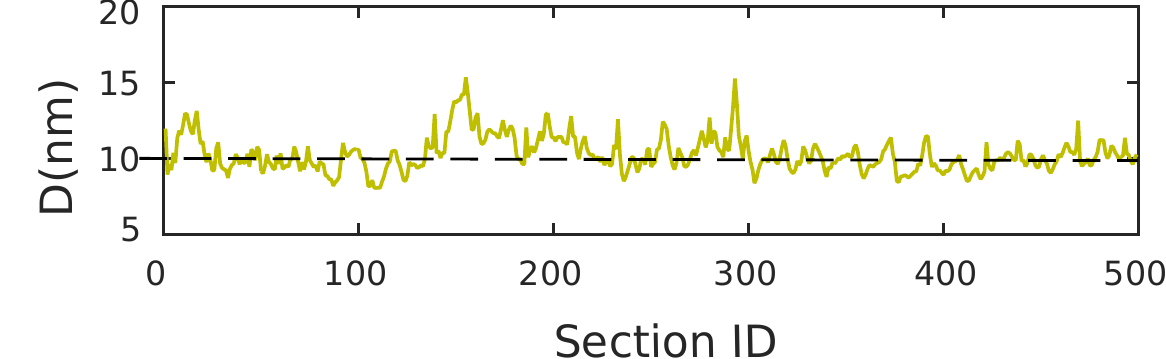}\label{fig:thicknessVariationFIBSEM}}
    \hfill
    \subfloat[]{\includegraphics[width=0.9\linewidth]{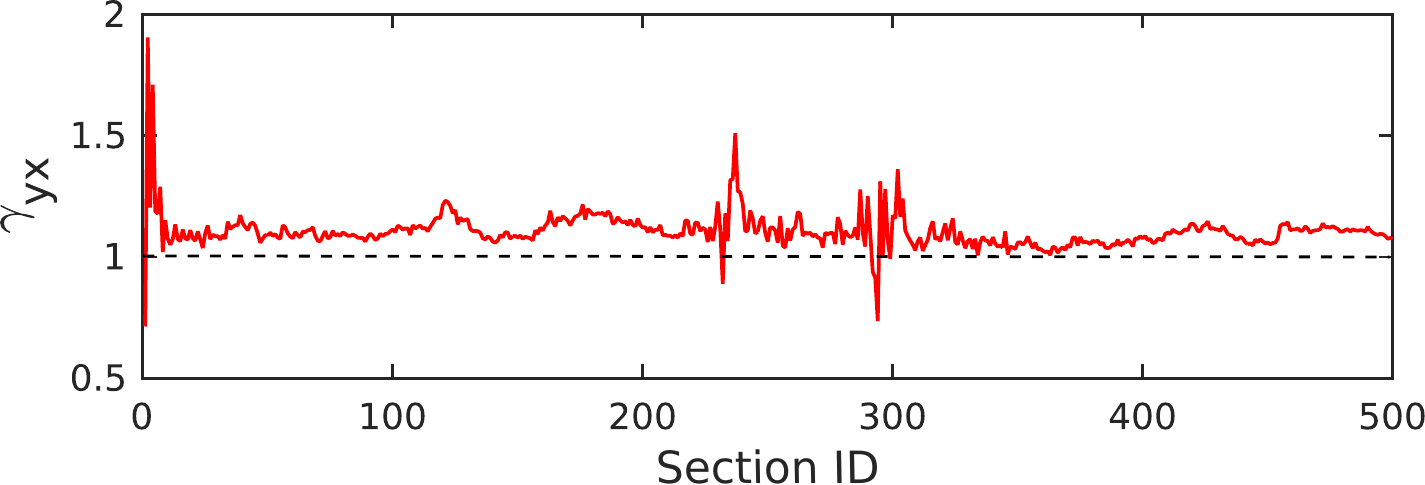}\label{fig:compressionVatiotionFIBSEM}}
  \caption{\small
  			\protect\subref{fig:thicknessVariationFIBSEM} Section thickness estimates (D) for a FIBSEM stack of 500 consecutive images (expected: $D \approx \SI{10}{nm}$)  
  			and \protect\subref{fig:compressionVatiotionFIBSEM} Estimates of the stretching coeff. $\gamma_{yx}$ for a FIBSEM stack of 490 sections.}
\label{fig:thickness_compression_figures}
\end{figure}
To validate the estimation of the stretching coefficient $\gamma$, we used linearly compressed versions of a synthetic image as shown in Figure~\ref{subfig:radial_grad}. The original image was composed of bright circular objects with radial gradients. Then the image was re-scaled with known $\gamma$ along the $y$ axis (vertical) down to different sizes. Using equation \eqref{eq:gamma} we recovered $\gamma$ with an average accuracy of $97.3\%$ for a linear compression of $75\%$ (Table \ref{tb:compression_all}).  

Estimated $\gamma$ for real data sets (ssTEM~\cite{ISBI2012} and FIBSEM) are summarized in Table \ref{tb:compression_all}. The FIBSEM dataset of 490 images was taken from songbird brain tissue imaged at $\SI{5}{nm} \times \SI{5}{nm}$ resolution along the $XY$ plane and with expected section thickness of $\SI{10}{nm}$. The entire FIBSEM stack had the dimensions $\SI{8}{\micro\metre} \times \SI{8}{\micro\metre} \times \SI{5}{\micro\metre}$. Estimates of $\gamma$ and section thickness for the FIBSEM dataset are plotted in Figure~\ref{fig:thickness_compression_figures}. 

To validate our thickness estimation method, we prepared a dataset of 20 serial sections, taken from the same brain area. 
Three image stacks were obtained by performing ssSEM on three non-overlapping areas of these sections.
The image size of each of these subvolumes are: (1) $\SI{9.5}{}{} \times \SI{9.5}{\micro\metre}$ 
(2) $\SI{6.5}{}{} \times \SI{6.5}{\micro\metre}$ 
(3) $\SI{6.5}{}{} \times \SI{6.5}{\micro\metre}$. 
The EM images were acquired at a spatial resolution of $\SI{5}{\nano\metre} \times \SI{5}{\nano\metre}$. 

We found that the FIBSEM data were associated with a higher $\gamma$ (lower linear compression) compared to the ssSEM data. This is due to the fact that unlike in ssSEM, FIBSEM does not make use of a diamond knife for thin sectioning, which is a potential source of linear compression. Instead, FIBSEM uses an ion beam to successively burn away thin layers.  

The image similarity measures used in our approach and Sporring \etal~\cite{sporring2014} are based on local deviations of pixel intensities across adjacent sections. Therefore, thickness estimates are sensitive to errors in image registration. For this reason, it is important to make sure that the image stack is properly registered before applying either of these methods for section thickness estimation.

We registered the serial section images into a 3D image volume using elastic alignment \cite{saalfeld2012} that jointly performs 2D stitching, 3D alignment, and deformation correction. This approach is based on an initial alignment obtained by matching image landmarks on nearby sections, where the landmarks are defined using SIFT image features. Further deformations are estimated using local block matching. Afterwards, this initial alignment is optimized by modeling each section as a mesh of springs where parts of the image are allowed to translate and rotate subject to imposed rigidity limits.

\begin{figure}[th!]
 	\centering
    \includegraphics[width=\linewidth]{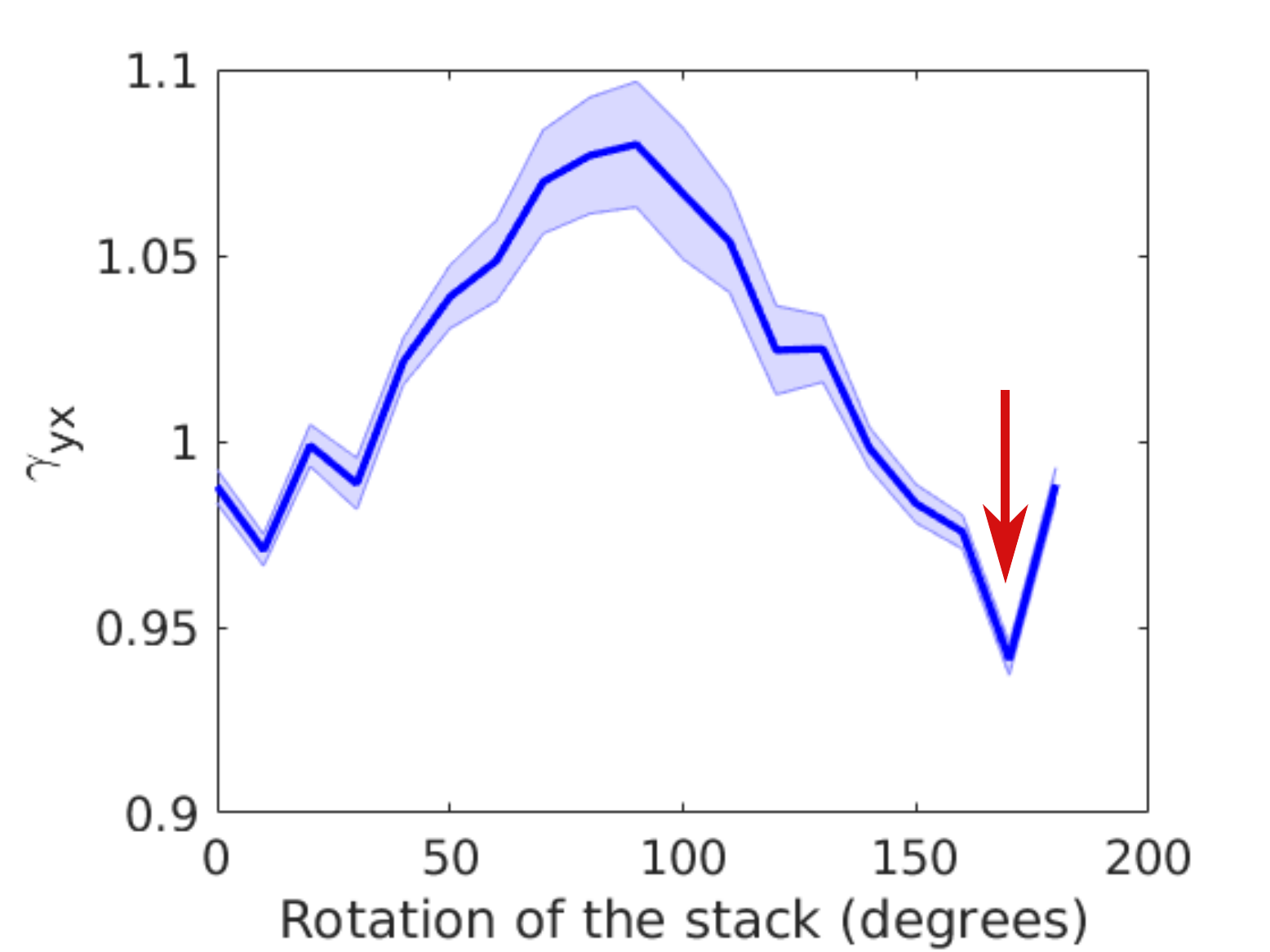}
  \caption{\small $\gamma_{yx}$ estimated for different rotations about the $Z$ axis. The minimum $\gamma_{yx}^* = 0.94$ corresponds to the stretching coefficient achieved at $170\degree$ rotation as pointed out by the arrow. 
  }
\label{fig:rotatedXY_stretching}
\end{figure}

As mentioned in Section~\ref{sec:compression}, the image axes $X$ and $Y$ are arbitrarily chosen. Therefore, we estimated the maximum stretching factor $\gamma_{yx}^*$ for a range of possible axes by rotating the original images up to $180\degree$.  Anisotropy estimates for a range of such rotations are shown in Figure~\ref{fig:rotatedXY_stretching}. We found that thickness estimation is optimal when the images are rotated such that the stretching factor $\gamma_{yx}$ is minimized. At this rotation angle, the $X$ axis is minimally stretched compared to the $Y$ axis. 

For the validation dataset, the average section thickness measured using AFM was $74.35 \pm \SI{2.64}{\nano\metre}$. Our method was able to estimate the thickness of the 20 sections with a mean absolute error of $9.91 \% \pm 1.97$, whereas the XY-averaging method in \cite{sporring2014} produced thickness estimates with a mean absolute error of $18.26 \% \pm 1.04$. The full comparison of estimation errors is plotted in Figures \ref{fig:afm_summary} and \ref{fig:AFMresults}. Because our estimation method is purely based on image statistics, it is prone to overestimating the thickness when images are noisy. An instance of such an overestimation is illustrated in Figure~\ref{fig:EM_explanation}.  

In addition to thickness measurements of ssSEM sections using AFM, we propose a second approach for validating thickness estimates. This second approach uses synthetic sections with known thicknesses derived from nearly isotropic FIBSEM volumes with known $XY$ resolution. We use the method described in Section~\ref{sec:thickness} to generate data points for learning the function given in Equation \eqref{eq:fs}. To validate section thickness estimates, we split each image stack into separate training and test data sets. The training sets were used to learn the regression function given by Equation~\eqref{eq:fs} and the test images were used for validation. We trained a regression function on 100 images of size $\SI{7}{\micro\metre} \times \SI{4.5}{\micro\metre}$ from a FIBSEM image stack, Figure~\ref{fig:distance_dissimilarity_curve}. We used the test images to create 3 separate image sequences of 30 images each with known displacements of $\SI{10}{nm}$, $\SI{50}{nm}$, and $\SI{75}{nm}$ along the relatively uncompressed axis. The results obtained are summarized in Table~\ref{tb:thicknessEstimates} along with a comparison with \cite{sporring2014}. 

Although included for comparison, we note that in \cite{sporring2014} an average distance-dissimilarity curve is generated for each pair of images between which the distance is estimated and therefore the interpolation function is based on the statistics of the validation data itself, unlike in our approach. 
A recent contribution towards correcting $Z$ coordinates of a 3D image stack is presented in \cite{hanslovsky2015} and \cite{hanslovsky2017}, in which relative $Z$ positions for each image is calculated. In order to compare with \cite{hanslovsky2015}, we converted absolute thickness estimates of our method and thickness measurements from AFM (subvolume 1 of the validation dataset) into relative thickness by normalizing the thickness values using the mean absolute thickness. With respect to relative thickness values obtained by AFM, our approach resulted in a mean absolute error of $0.13 \pm 0.03$ where as \cite{hanslovsky2015} resulted in a mean absolute error of $0.27 \pm 0.04$ (Figure~\ref{fig:relativeThicknessComparison}). For this comparison we used the Fiji~\cite{fiji2012} plugin available for \cite{hanslovsky2015} using its default parameters with the option to allow reordering disabled.


%
\begin{table}
\caption{\small Average thickness estimates for sets of 30 sections. The ``ground truth'' thicknesses were derived from nearly isotropic FIBSEM data as described in Section~\ref{sec:thickness}.}
\centering
\begin{tabularx}{\columnwidth}{@{}l XXX @{}}
\multicolumn{4}{c}{\small Thickness values are in nanometers (nm)} \\
\toprule[2pt]
\textbf{``Ground Truth''}& \textbf{10}	& \textbf{50} & \textbf{75}	\\
$xy$ avg. \cite{sporring2014}	& $9.93$		 & $47.35$			& $69.09$  			\\
\textit{Ours}		& $10.18 \pm 5.61$ & $47.02 \pm 5.60$ & $71.36 \pm 5.59$ \\
\bottomrule[2pt]
\end{tabularx}
\label{tb:thicknessEstimates}
\end{table}


\begin{table}[t!]
\caption{\small Estimated stretching coefficient $\gamma_{yx}$ in synthetic images (Figure~\ref{fig:squashed_images_curves}), 500 FIBSEM images, and 20 ssSEM images.}
\centering
\begin{tabularx}{1\columnwidth}{@{}l XX @{}l XXX}
\toprule[2pt]
\multicolumn{3}{l}{\textbf{$\gamma$ for synthetic images (Figure~\ref{fig:squashed_images_curves})}}
& \multicolumn{4}{l}{\textbf{$\gamma$ for real images}} \tabularnewline
\textit{Ground-Truth} 	& \textit{0.75} & \textit{0.50} & FIBSEM & \multicolumn{3}{l}{$0.94 \pm 6.6 \times 10^{-4}$}\\
Estimates 				& $0.73$		& $0.63$   & ssSEM & \multicolumn{3}{l}{$0.86 \pm 0.01$}\\
\bottomrule[2pt]
\end{tabularx}
\label{tb:compression_all}
\end{table}
\section{Conclusion}
\label{sec:conclusion}
We have presented a method for estimating both thickness and stretching in EM imagery, using image statistics alone. Our method is based on learning the distance between adjacent sections as a function of their dissimilarity.

The stretching coefficient quantifies the cumulative effect of different sources of anisotropy along the $XY$ plane including handling, storing, cutting, imaging, and the intrinsic anisotropy of the specimen. Anisotropy estimation is a useful pre-processing step for any method that assumes isotropy in image statistics. 

As part of this work, we have created a dataset of 20 ssSEM images along with thickness measurements directly obtained with AFM. We used this dataset to compare the performance of our thickness estimation method with other methods that use image statistics for indirect estimation of section thickness.

Thickness estimation methods based on image statistics alone are prone to be inaccurate if sample anisotropy is not taken into account. We have shown that estimation of  $XY$ anisotropy can help to improve the accuracy of thickness estimation. Our anisotropy estimation method selects the optimal rotation of the original image stack to train a regressor that is minimally affected by sample anisotropy. We recommend using images larger than $\SI{7}{} \times \SI{7}{\micro\metre}$ so that effects of locally oriented structures may even out given a larger scope.

\addtolength{\textheight}{-14cm}   
\section*{\normalsize{Acknowledgment}}
We thank Ziqiang Huang, for FIBSEM and ssSEM data, and Thomas Templier for ssSEM data.


%









\bibliographystyle{IEEEtran}
\bibliography{refs}
\end{document}